\documentclass[preprint,showpacs,preprintnumbers,amsmath,amssymb]{revtex4}

\usepackage{graphicx}
\usepackage{dcolumn}
\usepackage{bm}

\begin{document}


\title{Quark-Hadron Duality in Neutron (${\rm ^3}$He) Spin Structure}

\author{P.~Solvignon$^{21,1}$, N.~Liyanage$^{22}$, J.-P.~Chen$^{12}$, Seonho~Choi$^{21}$, 
K.~Aniol$^3$, T.~Averett$^{23}$, W.~Boeglin$^7$, A.~Camsonne$^2$, G.D.~Cates$^{22}$, 
C.C.~Chang$^{15}$, E.~Chudakov$^{12}$, B.~Craver$^{22}$, F.~Cusanno$^{9}$, A.~Deur$^{22}$, 
D.~Dutta$^4$, R.~Ent$^{12}$, R.~Feuerbach$^{12}$, S.~Frullani$^{9}$, H.~Gao$^4$, 
F.~Garibaldi$^{9}$, R.~Gilman$^{12,19}$, C.~Glashausser$^{19}$, V.~Gorbenko$^{11}$, O.~Hansen$^{12}$, 
D.W.~Higinbotham$^{12}$, H.~Ibrahim$^{18}$, X.~Jiang$^{19}$, M.~Jones$^{12}$, A.~Kelleher$^{23}$, 
J.~Kelly$^{15}$, C.~Keppel$^{8,12}$, W.~Kim$^{14}$, W.~Korsch$^{13}$, K.~Kramer$^{23}$, 
G.~Kumbartzki$^{18}$, J.J.~LeRose$^{12}$, R.~Lindgren$^{22}$, B.~Ma$^{17}$, D.J.~Margaziotis$^3$, 
P.~Markowitz$^7$, K.~McCormick$^{18}$, Z.-E.~Meziani$^{20}$, R.~Michaels$^{12}$, 
B.~Moffit$^{23}$, P.~Monaghan$^{17}$, C.~Munoz Camacho$^5$, K.~Paschke$^{16}$, B.~Reitz$^{12}$, 
A.~Saha$^{12}$, R.~Sheyor$^{20}$, J.~Singh$^{22}$, K.~Slifer$^{21}$, V.~Sulkosky$^{23}$, 
A.~Tobias$^{22}$, G.M.~Urciuoli$^{10}$, K.~Wang$^{22}$, K.~Wijesooriya$^4$, 
B.~Wojtsekhowski$^{12}$, S.~Woo$^{14}$, J.-C.~Yang$^6$, X.~Zheng$^1$, L.~Zhu$^{17}$ \\
~\\
(Jefferson Lab E01-012 Collaboration) \\
~}

\address{%
$^1$Argonne National Laboratory, Argonne, IL 60439\\
$^2$Universit\'{e} Blaise Pascal et CNRS/IN2P3 LPC, 63177 Aubi\`{e}re Cedex, France\\
$^3$California State University, Los Angeles, Los Angeles, CA 90032\\
$^4$Duke University, Durham, NC 27708\\
$^5$CEA Saclay, DAPNIA/SPhN, F-91191 Gif sur Yvette, France\\
$^6$Chungnam National University, Taejon 305-764, Korea\\
$^7$Florida International University, Miami, FL 33199\\
$^8$Hampton University, Hampton, Virginia 23187\\
$^{9}$Istituto Nazionale di Fisica Nucleare, Gruppo Collegato Sanit\'{a}, Seziona di Roma, 00161 Roma, Italy\\
$^{10}$Istituto Nazionale di Fisica Nucleare, Sezione di Roma, 00185 Roma, Italy\\
$^{11}$Kharkov Institute of Physics and Technology, Kharkov 61108, Ukraine \\
$^{12}$Thomas Jefferson National Accelerator Facility, Newport News, VA 23606\\
$^{13}$University of Kentucky, Lexington, KY 40506\\
$^{14}$Kyungpook National University, Taegu City, South Korea\\
$^{15}$University of Maryland, College Park, MD 20742\\
$^{16}$University of Massachusetts, Amherst, MA 01003\\
$^{17}$Massachusetts Institute of Technology, Cambridge, MA 02139\\
$^{18}$Old Dominion University, Norfolk, VA 23529\\
$^{19}$Rutgers, The State University of New Jersey, Piscataway, NJ 08855\\
$^{20}$University of Tel Aviv, Tel Aviv, 69978 Israel\\
$^{21}$Temple University, Philadelphia, PA 19122\\
$^{22}$University of Virginia, Charlottesville, VA 22904\\
$^{23}$College of William and Mary, Williamsburg, VA 23187
}

\date{\today}

\begin{abstract}
We present experimental results of the first high-precision test of quark-hadron duality in 
the spin-structure function $g_1$ of the neutron and $^3$He using a polarized $\rm{^3He}$ 
target in the four-momentum-transfer-squared range from 0.7 to 4.0 $(\mathrm{GeV}/c)^2$. 
Global duality is observed for the spin-structure function $g_1$ down to at least 
$Q^2 = 1.8~(\mathrm{GeV}/c)^2$ in both targets. We have also formed the photon-nucleon 
asymmetry $A_1$ in the resonance region for $^3$He and found no strong $Q^2$-dependence 
above 2.2 $(\mathrm{GeV}/c)^2$.
\end{abstract}

\pacs{13.60.Hb, 13.88.+e, 14.20.Dh}
\maketitle
Quark-hadron duality describes the remarkable similarity between electron-nucleon scattering 
in the deep inelastic scattering (DIS) region, where the electron scatters off an asymptotically 
free point-like quark, and in the nucleon resonance region where the electron scatters off a 
highly correlated cluster of quarks and gluons.

In high-energy electron-nucleon scattering, the electron interacts with the nucleon constituents 
by exchanging a virtual photon. This process depends on two variables, $Q^2$ and $x$. The 
quantity $Q^2$ is the four-momentum squared of the exchanged virtual photon and $x=Q^2/(2M\nu)$ 
is the Bjorken scaling variable, which represents the fraction of the nucleon four-momentum 
carried by the struck parton, with $M$ the mass of the nucleon and $\nu$ the energy transfer 
between the lepton and the nucleon. The invariant mass $W$ of the virtual photon-nucleon system 
is related to $Q^2$ and $x$ through $W^2 = M^2+Q^2/x-Q^2$. The scattering cross section is 
parametrized in terms of two unpolarized structure functions ($F_1$ and  $F_2$) and two polarized 
structure functions ($g_1$ and $g_2$). These structure functions are related to the polarized and 
unpolarized quark distribution functions in the nucleon. Another observable that provides a direct 
insight into the polarized quark distributions in the nucleon is the photon-nucleon asymmetry, 
$A_1$, which is approximately equal to the ratio of the polarized and unpolarized structure 
functions, $g_1/F_1$, at large $Q^2$.

In the DIS region, at asymptotically large values of $Q^2$ and $W$, the electron scatters off a 
point-like free quark. As a result, the structure functions are independent of $Q^2$ ($x$-scaling), 
modulo logarithmic corrections due to gluon radiation. On the other hand, at low values of $Q^2$ 
and $W$, the quarks strongly interact with each other and resonance production dominates. In this 
nucleon resonance region, the electron can be considered to be scattering off a correlated cluster 
of quarks and gluons, the nucleon. Thus, the structure functions are expected to depend strongly 
on $Q^2$. However, data taken at Stanford Linear Accelerator Center (SLAC)  more than 30 years ago 
revealed that the scaling curve seen  for DIS data at high $Q^2$ is an average over the resonances 
at lower $Q^2$ but at the same value of $x$~\cite{Bloom:1970xb}. Since then, there have been many 
experiments to study this intriguing observation and   numerous theoretical attempts to explain  
it. For a review of the full body of experimental and theoretical work see 
Ref.~\cite{Melnitchouk:2005zr}.

In the case of unpolarized electron scattering, duality can be expressed as 
\begin{equation}
\int_{x_1(W_1,Q^2)}^{x_2(W_2,Q^2)} dx~F_2^{\mathrm{res}}(x,Q^2) =  \int_{x_1}^{x_2} 
dx~F_2^{\mathrm{DIS}}(x,Q^2),
\label{eq:f2}
\end{equation}
where $F_2^{\mathrm{res}}(x,Q^2)$ is the structure function measured in the resonance region at 
low $Q^2$ while $F_2^{\mathrm{DIS}}(x,Q^2)$ is the scaling function evolved down to the same $Q^2$. 
Global duality is said to hold when Eq.~\ref{eq:f2}) is performed over the entire resonance region. 
Local duality would be observed if the above equality holds when integrated over a resonance 
bump~\cite{Close:2001ha,Close:2003wz}. Data from Jefferson Lab (JLab) Hall C~\cite{Niculescu:2000tj} have 
demonstrated that for unpolarized structure functions Bloom-Gilman duality holds at the 10\% level 
down to $Q^2$ = 0.5 $(\mathrm{GeV}/c)^2$. These data also show that local duality holds for each of 
the three prominent resonance regions.

The observation of duality in unpolarized structure functions raises the exciting question 
whether quark-hadron duality holds for spin-structure functions as well. Indeed, duality 
in spin-structure functions would be even more intriguing than in the unpolarized case, since 
the spin-structure functions come from the differences of cross sections, they are not 
necessarily positive definite. As a result, a spin-structure function measured at the same value 
of $x$ in resonance and DIS regions can, in principle, have opposite signs. For example, the 
contribution to $g_1^p$ at the $\Delta$  resonance is negative at low $Q^2$, while the scaling 
curve for  $g_1^p$ is positive.

Recent data from Jefferson Lab~\cite{Bosted:2006gp,Dharmawardane:2006zd} and 
DESY~\cite{Airapetian:2002rw} have shown that quark-hadron duality holds globally, in the 
case of polarized structure functions of the proton and the deuteron, down to $Q^2 = 1.7$ 
$(\mathrm{GeV}/c)^2$. However, the high-precision results from Jefferson Lab Hall B indicated 
that local duality is violated for the proton and deuteron polarized structure functions 
in the region of the $\Delta$  resonance even for $Q^2$ values as high as 
5.0 $(\mathrm{GeV}/c)^2$~\cite{Bosted:2006gp}.

Polarized structure functions measured previously over the low $Q^2$-range of 0.1-0.9 
$(\mathrm{GeV}/c)^2$, using a polarized $^3$He target as an effective neutron 
target~\cite{E94010}, have shown hints of quark-hadron duality for the polarized structure 
functions of the neutron. However, until now there have been no resonance-region neutron 
spin-structure data in the intermediate $Q^2$-range where quark-hadron duality is 
expected to manifest. 

In this letter we report results from Jefferson lab experiment 
E01-012 where we measured polarized structure functions in the resonance region over the 
$Q^2$-range 0.7-4.0 $(\mathrm{GeV}/c)^2$  using a polarized $^3$He target as an effective 
polarized neutron target. These new results provide a precision test of quark-hadron duality 
for neutron spin-structure functions. 


E01-012 took data in 2003 in Hall A at Jefferson Lab where inclusive scattering of 
longitudinally polarized electrons from a longitudinally or transversely polarized $\rm{^3}$He 
target was studied. We formed asymmetries and polarized cross section differences from data taken 
at a scattering angle of 25$^{\circ}$ for three incident beam energies, 3.028, 4.018 and 5.009 GeV, 
and at 32$^{\circ}$ for an incident beam energy of 5.009 GeV.

The polarized electron beam was produced from a strained GaAs photocathode which was 
illuminated by circularly polarized light. The helicity of the beam was pseudo-randomly
reversed at a rate of 30 Hz. The beam polarization was monitored by a M{\o}ller polarimeter 
and was found to be stable during each incident energy running but varied between 70 and 85\% 
for the production data, depending on the incident beam energy values and the status of the 
other experimental halls. The total relative uncertainty on the beam polarization was 3.4\%. 
The beam current delivered on the target was kept below 15 $\mu$A in order to limit target 
depolarization. 

The polarized target consists of two connected chambers with an admixture of rubidium (Rb) and 
$\rm{^3}$He in the upper chamber where spin-exchange optical pumping~\cite{RevModPhys.44.169} 
takes place. About 90 W of 795 nm light produced by fiber-coupled laser diode arrays and an oven 
heated to 170$^{\circ}$C were used to this goal. Under running conditions, the $\rm{^3}$He 
density was $3.20 \times 10^{20}$ cm$^{-3}$ in the lower chamber where the electron beam passed 
through and from which the Rb was absent due to the lower temperature. The polarization of the 
target was monitored by two independent polarimetries: one based on nuclear magnetic resonance 
(NMR)~\cite{Abragam} and the other on the electron paramagnetic resonance frequency shift 
(EPR)~\cite{PhysRevA.58.3004}. Both polarimetries rely on the adiabatic fast passage (AFP) 
technique~\cite{Abragam}. The average target polarization was 0.38 $\pm$ 0.02.  The cells also 
contain a small amount of nitrogen gas (at about 1\% of the $\rm{^3}$He density) to quench 
radiative transitions which limit the optical pumping efficiency. However, this introduces a 
dilution factor in the extracted physics quantities. Nitrogen dilution factors were determined 
from measurements on a reference cell filled with nitrogen gas and were found to dilute the 
asymmetries by 5 to 9\%. The asymmetry of elastic scattering $\vec{e}-\vec{\rm{^3He}}$ was 
extracted from data taken at an incident beam energy of 1.046 GeV and a scattering angle of 
16$^{\circ}$ in order to obtain an independent measurement of the product of beam and target 
polarizations and nitrogen dilution. Theoretical and experimental elastic asymmetries were found 
to agree within a total uncertainty of 5\%~\cite{PAT}.

Both Hall A high resolution spectrometers~\cite{Alcorn:2004sb} were set at the same momentum 
and angle configurations and tuned to optimize the detection of the scattered electrons. This 
symmetric configuration has the advantage of obtaining two independent measurements and 
therefore controlling part of our systematic uncertainties. A gas \v{C}erenkov counter 
combined with a two-layered lead-glass calorimeter provided particle identification with a pion 
rejection better than 10$^3$ to 1 while keeping the electron efficiency above 99\%. The 
remaining contamination from pions to the asymmetries has a maximum effect on the asymmetries 
of 15 ppm~\cite{PAT}, which is negligible compared to the statistical uncertainties. 

We formed the longitudinal and transverse polarized cross section differences using:
\begin{eqnarray}
\Delta\sigma_{\parallel(\perp)} = \frac{d^2 \sigma^{\downarrow \Uparrow (\Rightarrow)}}{ 
d\Omega dE^{\prime}} - \frac{d^2 \sigma^{\uparrow \Uparrow (\Rightarrow)}}{d\Omega dE^{\prime}}
\label{eq:dsig}
\end{eqnarray}
where the superscript $\uparrow$ ($\downarrow$) represents the incident electron spin direction and 
$\Uparrow$ ($\Rightarrow$) the longitudinal (transverse) target spin direction. The quantities 
$E^{\prime}$ and $\Omega$ correspond 
to the scattered electron energy and the spectrometer solid angle, respectively. The 
polarized structure function $g_1$ can be formed directly from the polarized cross section 
differences as follows:
\begin{eqnarray}
g_1 = \frac{M Q^2 \nu}{4 \alpha_e^2} \frac{1}{E^{\prime} (E + E^{\prime})}
[\Delta\sigma_{\parallel} + \tan\frac{\theta}{2} \Delta\sigma_{\perp}] 
\label{eq:g1}
\end{eqnarray}
Here, $\alpha_e$ is the fine structure constant, $\theta$ is the scattering angle and 
$E$ is the energy of the incident electron beam. 


External and internal radiative corrections have been applied to the polarized 
cross section differences following the formalism of Mo and Tsai~\cite{MoTsai}
for the spin-independent part and the approach of Akushevich and 
Shumeiko~\cite{Akushevich:1994dn} for the spin-dependent part. 

\begin{figure}
\vspace{0.1cm}
\includegraphics[scale=0.55]{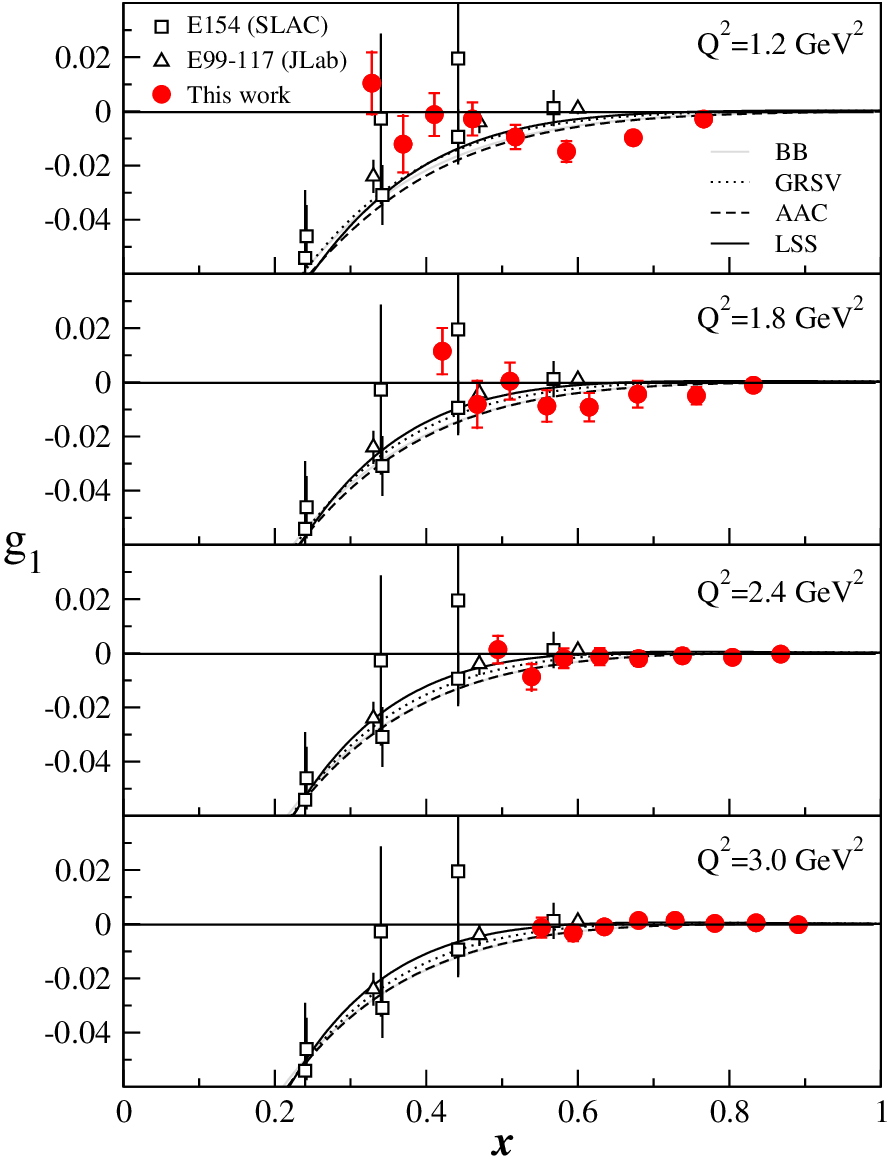}
\caption{\label{fig:g1}  The spin-structure function $g_1^{\rm ^3He}$ (per nucleon) 
in the resonance region at $Q^2$-values of 1.2, 1.8, 2.4 and 3.0 $(\mathrm{GeV}/c)^2$. 
The error bars represent the total uncertainties with the inner part being statistical 
only. Also plotted are the DIS world data from E154~\cite{Abe:1997cx} and 
E99-117~\cite{ZHENG} (note that these data are at different $Q^2$). The curves were 
generated from the NLO parton distribution functions of 
Ref.~\cite{Blumlein:2002be,Gluck:2000dy,Goto:1999by,Leader:2005kw} to which target-mass 
corrections were applied.}
\end{figure}

The structure function $g_1$ was generated at constant energies and scattering angles. 
Our goal in this analysis is the quantitative study of quark-hadron duality for the 
spin-structure function $g_1$. Following the same procedure as in the unpolarized 
case (see Eq.~\ref{eq:f2}), $g_1$ would be integrated at constant $Q^2$ over a limited $x$-range.
However forming this partial moment requires the interpolation of $g_1$ to a constant $Q^2$. 
The coverage of our data allows interpolation of $g_1$ to the four $Q^2$-values 
of 1.2, 1.8, 2.4 and 3.0 $(\mathrm{GeV}/c)^2$. The results for the $^3$He spin-structure function 
$g_1$ are shown in Fig.~\ref{fig:g1}. The DIS parameterizations of $g_1$ for the proton and the 
neutron, from Bl\"{u}mlein and B\"{o}ttcher (BB)~\cite{Blumlein:2002be}, GRSV group~\cite{Gluck:2000dy}, 
AAC collaboration~\cite{Goto:1999by} and Leader, Sidorov and Stamenov (LSS)~\cite{Leader:2005kw}, 
were evolved to those $Q^2$-values and then combined to obtain the $^3$He DIS parametrizations 
using the effective polarization equation~\cite{Bissey:2001cw}:
\begin{eqnarray}
g_1^{\rm ^3He} = p_n g_1^n + 2 p_p g_1^p.
\label{eq:g1he3ton}
\end{eqnarray}
The effective polarizations of the neutron and the proton are $p_n = 0.86 \pm 0.02$ and 
$p_p = -0.028 \pm 0.004$, respectively~\cite{Friar:1990vx}. While neglecting Fermi motion and 
nucleon binding effects (EMC effect), this method of computing the $^3$He structure function 
from the proton and the neutron structure functions was confirmed to be a good approximation at 
the 4\% level, which is well below the DIS parametrization precision. All parameterizations were 
taken at Next-to-Leading Order (NLO). Target-mass corrections~\cite{Sidorov:2006fi} have been 
applied to the DIS parameterizations in order to take into account the finite mass of the nucleon.
Also plotted are the highest precision world data on $g_1^{\rm ^3He}$ in the DIS region from 
SLAC E154~\cite{Abe:1997cx} and JLab E99-117~\cite{ZHENG}. It can be observed that 
our resonance data at $Q^2 = 1.2~(\mathrm{GeV}/c)^2$ oscillate around the DIS curves and
are significantly lower in the $\Delta$(1232) region ($x \simeq$ 0.65). As $Q^2$ increases, 
our data approach the DIS parameterizations indicating quark-hadron duality.


To perform a quantitative test of quark-hadron duality, $g_1$ was integrated over 
the $x$-interval corresponding to the resonance region~\cite{Bianchi:2003hi} covered 
by our data, which is from pion threshold ($x_{\pi}$) to an $x$ corresponding to 
$W=1.905$ GeV ($x_{min}$):
\begin{eqnarray}
\tilde{\Gamma}_1(Q^2) = \int_{x_{min}}^{x_{\pi}} dx~g_1(x,Q^2)
\label{eq:dual}
\end{eqnarray}

\begin{figure}
\vspace{0.1cm}
\includegraphics[scale=0.55]{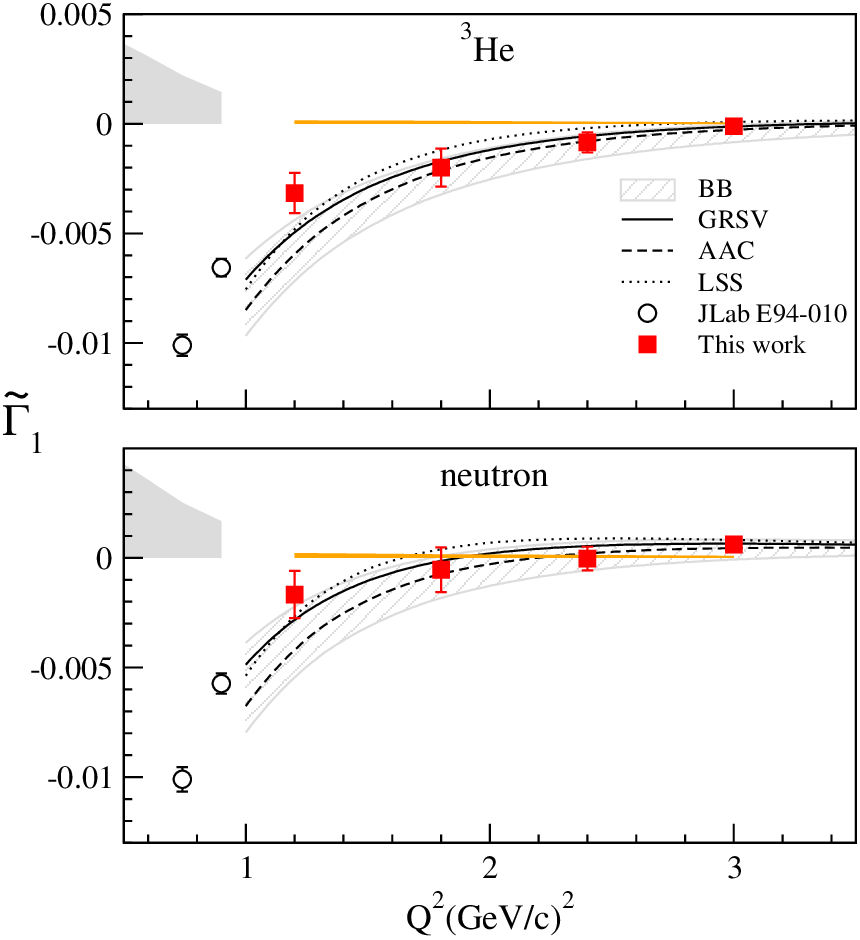}
\caption{\label{fig:gam1} $\tilde{\Gamma}_1^{\rm ^3He}$ and $\tilde{\Gamma}_1^n$: test 
of spin duality on ${\rm ^3He}$ (top) and neutron (bottom). Also plotted are the DIS 
parameterizations of Bl\"{u}mlein and B\"{o}ttcher~\cite{Blumlein:2002be} (grey band), 
GRSV~\cite{Gluck:2000dy} (solid curve), AAC~\cite{Goto:1999by} (dashed curve) and 
LSS~\cite{Leader:2005kw} (dotted curve) after applying target-mass corrections. The open 
circles are data from JLab E94-010~\cite{E94010}.
The bands on the $x$-axis represent the systematic uncertainty of each data set.}
\end{figure}

The experimental $g_1$-integral for the neutron was extracted using the method 
described in Ref.~\cite{CiofidegliAtti:1996cg}:
\begin{eqnarray}
\tilde{\Gamma}_1^n = \frac{1}{p_n}\tilde{\Gamma}_1^{\rm ^3He} 
                      - 2\frac{p_p}{p_n}\tilde{\Gamma}_1^p
\label{eq:he3ton}
\end{eqnarray}
Data from JLab Hall B experiment on the proton spin-structure function~\cite{Bosted:2006gp} 
were used for $\tilde{\Gamma}_1^p$. Figure~\ref{fig:gam1} shows the comparison of 
the integral of $g_1$ over the resonance region to the DIS parameterizations. Data from JLab 
E94-010~\cite{E94010} are also plotted and a smooth transition can be observed with our 
data.

At the lowest $Q^2$, our data deviate slightly from the DIS parameterizations for both $^3$He 
and the neutron. As $Q^2$ increases, the resonance data and the DIS parametrization show good 
agreement indicating global quark-hadron duality holds at least down to 
$Q^2$ = 1.8 $(\mathrm{GeV}/c)^2$ and above. It is important to note that, for the proton and 
the deuteron, global duality was previously observed for $Q^2$ above 1.7 
$(\mathrm{GeV}/c)^2$~\cite{Bosted:2006gp,Wesselmann:2006mw}.

Quark-hadron duality was also studied for the photon-nucleon asymmetry $A_1$. From 
the parallel and perpendicular asymmetries $A_1$ was extracted as follows:
\begin{eqnarray}
A_1(x,Q^2) =  \frac{A_{\parallel}(x,Q^2)}{D(1+\zeta \eta)} - \frac{\eta A_{\perp}(x,Q^2)}
{d(1+\zeta \eta)}
\label{photonasym1}
\end{eqnarray}
where  
\begin{eqnarray*}
D&=&\frac{1 - \epsilon \frac{E^{\prime}}{E}}{1+ \epsilon R(x,Q^2)}, 
~d = D \sqrt{\frac{2\epsilon}{1+\epsilon}},
~\zeta = \eta \frac{1+\epsilon}{2\epsilon} \\
\eta&=&\frac{\epsilon \sqrt{Q^2}}{E -E^{\prime} \epsilon},
~\epsilon = \frac{1}{1+2(1+\nu^2/Q^2)\tan^2\frac{\theta}{2}}
\label{coefa1a2}
\end{eqnarray*}
$D$ is the photon depolarization factor and $\epsilon$ is the photon polarization. The 
longitudinal to transverse cross section ratio $R(x,Q^2)$ has never been measured 
on ${\rm ^3He}$ in the resonance region. Precise data on this ratio exist for the proton 
and the deuteron in the DIS region~\cite{Tvaskis:2006tv} and show a reduction of deuteron ratio
by 30\% compared to the proton, i.e. $R_D = 0.7 R_p$. Considering that quark-hadron duality 
holds for $R$~\cite{Melnitchouk:2005zr}, the same reduction of the ratio $R$ can be expected 
in the resonance region. Besides the ratio $R$ is believed to diminish further for $^3$He.
Therefore, it was estimated for $^3$He that $R = 0.5 R_p$ with $R_p$ extracted from fits to the
resonance data~\cite{Liang:2004tj}. We then studied the sensitivity of $A_1$ to different models 
and assumptions for $R$. The systematic effects on $A_1$ were found to be significantly smaller 
than the relative statistical uncertainties and were added to the systematic uncertainties.

\begin{figure}
\vspace{0.12cm}
\includegraphics[scale=0.4]{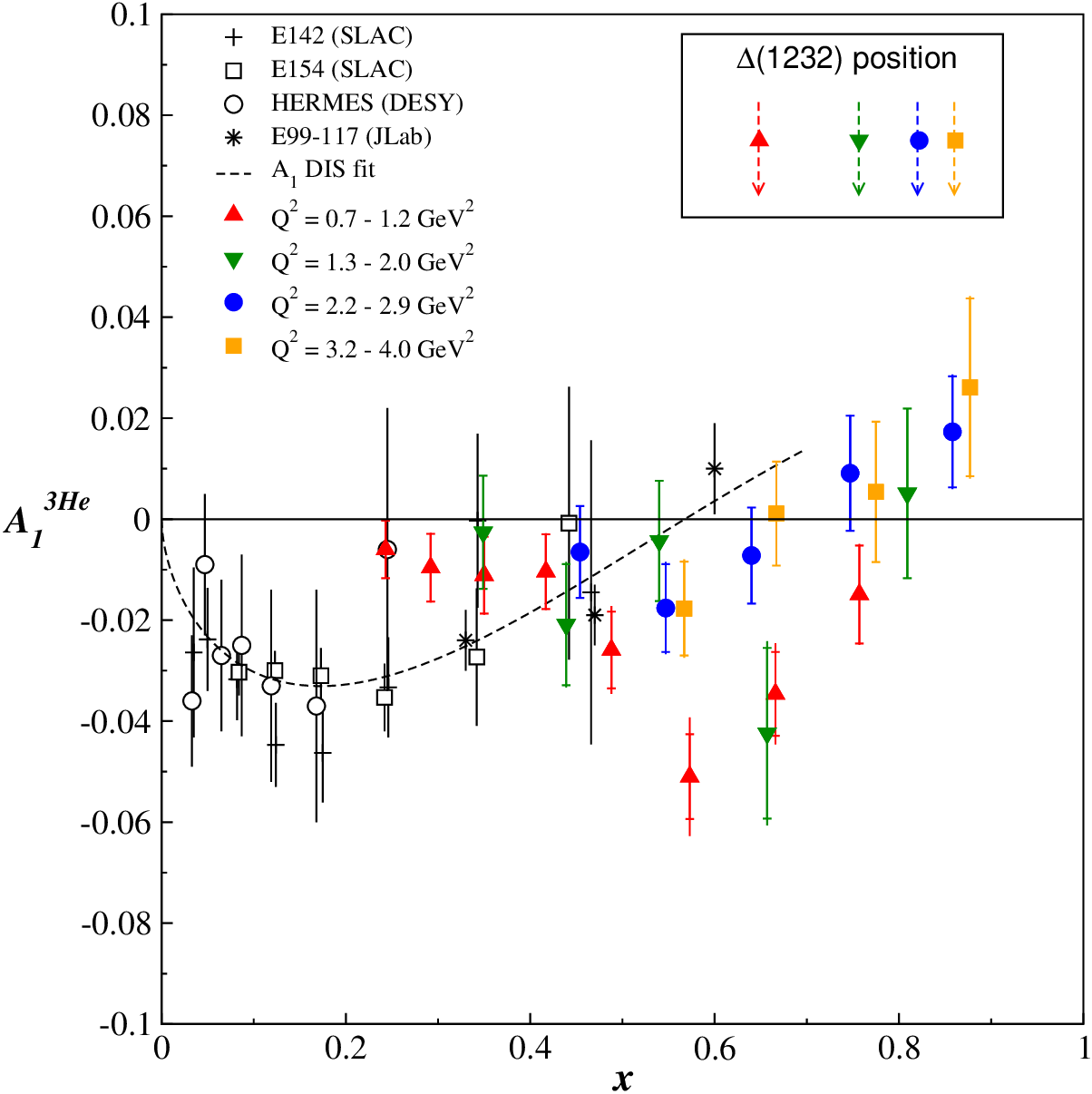}%
\caption{\label{fig:A1he} $A_1^{\rm ^3He}$ in the resonance region. DIS data are from SLAC  
E142~\cite{Anthony:1996mw}, E154~\cite{Abe:1997cx}, from DESY experiment 
HERMES~\cite{Ackerstaff:1999ey} and from JLab E99-117~\cite{ZHENG}. The error bars 
represent the total uncertainties with the inner part being statistical only. The curve 
represents a fit to the $A_1^{\rm ^3He}$ DIS data (see text). The arrows in the black frame 
point to the $\Delta$(1232) peak position for each of our data sets}
\end{figure}

The photon-nucleon asymmetry $A_1$ for $^3$He in the resonance region is presented in 
Fig.~\ref{fig:A1he}. Also plotted are the DIS data in order to provide a direct 
comparison between the deep inelastic scattering behavior and the resonance data 
trend. We performed a parametric fit to the $^3$He DIS data resulting in: 
$x^{0.580} (-0.135 + 0.336 x -0.174 x^2)(1 + 0.578/Q^2)$. For $Q^2$ below 2.0 $(\mathrm{GeV}/c)^2$,
it can be seen that $A_1^{\rm ^3He}$ in the vicinity of the $\Delta$(1232) peak is 
large and negative unlike the DIS behavior. At higher $Q^2$, our $A_1^{\rm ^3He}$ 
data cross zero and become positive even in the $\Delta$(1232) region. This is due to 
the increasing importance of the non-resonant background with respect to the resonance 
strength. The most noticeable feature is that, for values of $Q^2$ above 
2.2 $(\mathrm{GeV}/c)^2$, $A_1^{\rm ^3He}$ from our two data sets agrees very well 
indicating little or no $Q^2$-dependence. Furthermore, the trend of these data show 
the same particularity as the DIS data from E99-117 of becoming positive at high 
$x$, a prediction for DIS data from both pQCD inspired models~\cite{Brodsky:1994kg} and 
Relativistic Constituent Quark models~\cite{Isgur:1998yb}.

In summary, the results presented in this paper are the first precision test of quark-hadron 
duality for the neutron and $^3$He spin-structure functions. Global duality has been demonstrated 
to hold for the neutron and $^3$He polarized structure function $g_1$ down to at least $Q^2$ = 1.8 
$(\mathrm{GeV}/c)^2$. We also observed no strong $Q^2$-dependence in the photon-nucleon 
asymmetry $A_1^{\rm ^3He}$ measured in the resonance region for $Q^2 \ge 2.2~(\mathrm{GeV}/c)^2$
 and that $A_1^{\rm ^3He}$ becomes positive at large $x$.

We would like to acknowledge the outstanding support from the Jefferson Lab Hall A 
technical staff. This work was supported in part by the National Science Foundation and 
the US Department of Energy (DOE) Contract No. DE-AC05-84ER40150 Modification No. M175,
 under which the Southeastern Universities Research Association (SURA) operates the Thomas 
Jefferson National Accelerator Facility.


\end{document}